\documentclass[superscriptaddress,aps,preprintnumbers,amsmath,showpacs,amssymb,prd,nofootinbib,preprint]{revtex4-1}
\pdfoutput=1
\usepackage{booktabs} 
\usepackage{array} 
\usepackage[table,xcdraw]{xcolor} 
\usepackage{amsmath,amssymb}
\usepackage{graphicx}
\usepackage{fancyhdr}
\usepackage{bm, color}
\usepackage{slashed,amsthm,amsfonts,empheq}
\usepackage[caption=false]{subfig}
\usepackage{hyperref}
\usepackage{booktabs}
\usepackage{multirow}
\usepackage[left=2cm,right=2cm,top=2cm,bottom=2cm,includefoot,a4paper]{geometry}
\usepackage{tikz}
\usetikzlibrary{arrows.meta}

\newcommand{\Slash}[1]{{\ooalign{\hfil#1\hfil\crcr\raise.167ex\hbox{/}}}}

\newcommand{\beq}{\begin{equation}}  \newcommand{\eeq}{\end{equation}}
\newcommand{\bef}{\begin{figure}}  \newcommand{\eef}{\end{figure}}
\newcommand{\bec}{\begin{center}}  \newcommand{\eec}{\end{center}}

\newcommand{\laq}[1]{\label{eq:#1}}  
\newcommand{\Eq}[1]{Eq.(\ref{eq:#1})}

\newcommand{\vev}[1]{\left\langle {#1} \right\rangle}

\def\({\left(}
\def\){\right)}

\def\O{\mathcal{O}}

\newcommand{\AND}{~{\rm and}~}
\newcommand{\EV}{\,{\rm eV}}
\newcommand{\KEV}{\,{\rm keV}}

\newcommand{\GEV}{\,{\rm GeV}}
\newcommand{\TEV}{\,{\rm TeV}}

\def\f{\phi}
\def\g{\gamma}

\def\l{\lambda}

\def\w{\omega}
\def\x{\xi}
\def\y{\eta}

\def\*{\dagger}

\begin{document}

\title{Constant Scaling Fails for Global Monopole Networks}

\author{Wakutaka Nakano}
\affiliation{Department of Physics, Tokyo Metropolitan University, Minami-Osawa, Hachioji-shi, Tokyo 192-0397, Japan}

\author{Wen Yin}
\affiliation{Department of Physics, Tokyo Metropolitan University, Minami-Osawa, Hachioji-shi, Tokyo 192-0397, Japan}

\begin{abstract}
Global monopoles, which can also be understood as the zero-gauge-coupling limit
of gauged monopoles, can form in the early Universe and evolve following
a scaling network, as do other topological defects.  However, only a limited
number of numerical studies have investigated their scaling behavior.  In this
Letter, we show that the monopole number density parameter, $\xi$, does not
follow the commonly assumed constant scaling.  Instead, it exhibits a 
logarithmic-like evolution, despite the fact that the energy of an isolated
global monopole grows linearly, rather than logarithmically, with the infrared
cutoff.  This behavior is found using the fat-monopole prescription and is
supported by a conventional fixed-core simulation.  We characterize the
deviation by the fractional response
$\gamma\equiv d\log \xi/d\log(m_r/H)$, and find
$\gamma\sim 0.5$ for blue-tilted initial spectra.  These results suggest that
analytic studies based on the assumption of constant scaling should be
revisited.  They are also relevant to cosmological scenarios involving monopole
dark matter, axion and dark-photon dark matter produced by monopole networks,
monopole-induced primordial black holes, and gravitational-wave production from
monopole dynamics.
\end{abstract}
\maketitle

\section{Introduction}

Topological defects are a generic prediction of theories with spontaneous
symmetry breaking in the early Universe~\cite{Kibble:1976sj,Vilenkin:2000jqa}.
Among them, global monopoles arise when a global $O(3)$ symmetry is broken to
$O(2)$, leaving a vacuum manifold with a nontrivial second homotopy group,
$\pi_2(S^2)=\mathbb{Z}$~\cite{Barriola:1989hx}.  The corresponding gauged
theory contains the 't Hooft--Polyakov monopole~\cite{tHooft:1974kcl,Polyakov:1974ek};
the simulations below study the global, or equivalently weak-gauge-coupling,
limit in which the long-range Goldstone field dominates the monopole energy.
Monopole--antimonopole pairs experience an approximately constant attractive
force, leading to efficient annihilation and the emergence of a scaling regime.

The cosmological evolution of global monopole networks has been investigated in
a series of numerical studies~\cite{Bennett:1990xy,Yamaguchi:2001rf,Yamaguchi:2001xn},
which found that the network rapidly approaches a scaling solution.  In this
regime, the monopole number density scales as
$n_M\propto t^{-3}$, corresponding to an approximately constant number of
monopoles within each Hubble volume.  This constant-scaling solution has become
the standard picture of global-monopole evolution and forms the basis of many
cosmological applications involving global monopoles.

The same picture has also been used as the basis for analytic and semi-analytic
descriptions of global-monopole dynamics.  In particular, velocity-dependent
one-scale descriptions assume that the network can be characterized by a single
length scale, or equivalently by an approximately constant number of monopoles
per Hubble volume, together with a root-mean-square velocity
\cite{Martins:2008ks,Sousa:2017wvx}.  Related simulations and CMB analyses also
identify a scaling regime and use it to calibrate effective network parameters
or to extrapolate stress-energy correlators over cosmological time scales
\cite{Lopez-Eiguren:2016jsy,Lopez-Eiguren:2017dmc}.  Thus, the scaling parameter has not merely been a numerical diagnostic; it has
been an input to the standard dynamical and cosmological description of
global-monopole networks. It is therefore essential to understand whether departures from constant scaling exist.

Recently, considerable attention has been paid to the precision of defect
scaling laws.  In particular, numerical simulations of global cosmic strings
have shown that the scaling solution receives logarithmic corrections
associated with the infrared sensitivity of the Goldstone field
\cite{Hiramatsu:2010yu,Gorghetto:2018myk,Gorghetto:2020qws,Saikawa:2024bta,Kim:2024wku,Buschmann:2024bfj,Benabou:2024msj,Kim:2024dtq}
(see also \cite{Sikivie:1982qv,Vilenkin:1982ks,Davis:1986xc,Harari:1987ht}
for earlier work and \cite{Dine:2020pds,Hindmarsh:2021zkt,Saikawa:2024bta}
for discussions).  These logarithmic corrections have important consequences
for dark matter axion production and have substantially changed quantitative
predictions for axion cosmology.\footnote{
A precise understanding of the scaling behavior may also help distinguish the
feebly coupled Peccei--Quinn scenario from the conventional post-inflationary
axion scenario.  The former corresponds to an alternative parameter region in
which the quality and hierarchy problems are alleviated and fat strings,
relevant for axion production, are predicted~\cite{Yin:2024txg,Yin:2024pri}.
}

It has been argued that the apparent logarithmic growth may instead be
interpreted as a transient relaxation toward a constant-scaling attractor,
with the available dynamical range being insufficient to reach the fixed
point.    Recent studies, mostly of domain wall
networks, another class of topological defects, have shown that the scaling solution fails to be a local attractor and can
remain sensitive to superhorizon long-range correlations
\cite{Gonzalez:2022mcx,Kitajima:2023kzu,Yin:2024pri}.  The physical reason is
simple.  If long-range correlations are generated, for example during
inflation, causality forbids subhorizon network evolution from rearranging the
field configuration on superhorizon scales.  The existence of such logarithmic corrections in global string system
naturally raises the question of whether other global defects exhibit analogous
deviations from the na\"{i}ve scaling solution since global monopoles are also
infrared-sensitive objects.

In this Letter, we revisit the scaling evolution of global monopoles using
lattice simulations over a substantially larger dynamical range than in
previous studies.  We find that, across a broad set of initial spectra and
simulation parameters, the monopole network does not approach the commonly
assumed constant-scaling solution.  Instead, the monopole number density
exhibits a positive logarithmic correction with respect to the hierarchy
between the radial-mode mass scale and the Hubble scale.  This behavior is
nontrivial because the energy of an isolated global monopole grows linearly,
rather than logarithmically, with the infrared scale.  We argue that the
relevant logarithm is associated not with the isolated-monopole energy, but
with the time required for monopole--antimonopole cores to collide and
annihilate.  Since the typical separation is of order the Hubble length and
the core size is set by the inverse radial-mode mass, this annihilation time
can depend on the hierarchy between these two scales.  We measure this running
scaling quantitatively, determine its dependence on the cosmological
background, and discuss its cosmological implications.

The existence of scaling corrections is relevant for any
cosmological scenario involving global monopoles, or local monopoles with a
sufficiently weak gauge coupling.  Since the monopole abundance controls dark
matter production, annihilation rates, energy injection, gravitational-wave
production~\cite{Aburatani:2026rct}, and primordial black hole formation~\cite{Aburatani:2026rct}, even a slowly varying
correction can lead to appreciable effects when extrapolated over the enormous
hierarchy between the Higgs mass scale and cosmological timescales.  In
addition, analytical descriptions of monopole-network evolution should take
corrections to constant scaling into account.

\section{Global monopole scaling and lattice setup}
\subsection{Global monopoles and their properties}

We briefly review the aspects of global monopoles relevant for their network
evolution.  Global monopoles arise when a global $O(3)$ symmetry is
spontaneously broken to $O(2)$, leaving a vacuum manifold
$S^2$ with $\pi_2(S^2)=\mathbb{Z}$ \cite{Barriola:1989hx,Vilenkin:2000jqa}.  A
minimal realization is given by
\begin{equation}
\mathcal{L}
=
\frac12\partial_\mu\phi^a\partial^\mu\phi^a
-
V, ~\quad V=\frac{\lambda}{4}
\left(\phi^a\phi^a-v^2\right)^2,
\end{equation}
with $a=1,2,3$.

Unlike local monopoles, the energy of a global monopole is dominated by the
Goldstone gradients outside the core.  Consequently, the energy contained
within a sphere of radius $R$ grows linearly,
\begin{equation}
E(R)\simeq 4\pi v^2R,
\end{equation}
where the infrared cutoff is determined by the distance to the nearest antimonopole.
This infrared-sensitive energy is the key property governing the dynamics of
global monopoles.

Since the energy increases with monopole--antimonopole separation, a pair
experiences an approximately constant attractive force,
\begin{equation}
F\simeq 4\pi v^2,
\end{equation}
which efficiently drives monopole--antimonopole annihilation after the phase
transition.  As a result, monopoles within each causal region continue to
annihilate until only an $\mathcal{O}(1)$ net winding remains per Hubble volume,
while monopoles separated by super-horizon distances cannot annihilate
causally. This provides a physical explanation for why the monopole abundance is expected
to become of order one per Hubble volume, but it does not by itself imply a
strictly constant scaling parameter, for example because the annihilation time
may depend on the relevant scale hierarchy.

In the following we use dimensionless variables
\begin{align}
  \hat t &= m_0 t, &
  \hat x &= m_0 x, &
  \hat\f_a &= \frac{\f_a}{v}, \nonumber\\
  \hat\l &= \lambda \frac{v^2}{m_0^2}, &
  \hat V &= \frac{\hat\lambda}{4}
  \left(\hat\f^a\hat\f_a-1\right)^2 .
\end{align}
Here $m_0$ is a reference mass scale.  After this rescaling the equation of
motion has the same form in terms of the hatted variables, so different
choices of $m_0$ and $v$ can be restored by dimensional analysis.  In the
following we drop the hats for notational simplicity.

\subsection{Setup for the simulation}

\paragraph{Lattice Model}
We perform lattice simulations to check the scaling behavior of the global network by modifying 3D {\tt CosmoLattice}~\cite{Figueroa:2020rrl,Figueroa:2021yhd} to solve the equation of motion 
\beq
\ddot{\f}_a -\frac{\Delta}{a^2}{\f}_a+ 3H \dot{\f }_a=- \partial_{\f^a}V 
\eeq
Here $H$ is the Hubble parameter, dots denote derivatives with respect to time, and $\Delta$ is the spatial Laplacian in comoving coordinates. 
We use $N_{\rm lat}=256$ lattice cells per side.  

\paragraph{Initial conditions}
At the initial time, $\eta=1/m_0$ with initial Hubble parameter $H_0=0.5m_0$, 
we consider the initial power spectrum of the scalar field as 
\beq 
\quad {\cal P}_{a,b}= \frac{1}{4}\left(\frac{k}{k_{\rm UV}}\right)^{p} k_{\rm UV}^{-1} \delta_{a,b}
\eeq 
where $k_{\rm UV}$ is also the UV cutoff of the initialized fluctuation and $\vev{\phi_{a,\vec k}\phi_{b,\vec{k}'}}=(2\pi)^3 \delta^{3}(\vec k+\vec{k}'){\cal P}_{a,b}(k)$, with $\vev{}$ being the ensemble average. The averages of the fields are initially in the symmetric phase, $\vev{\f_a}=0$.
 We use
$p=\{-3,-2,-1,0,1\}$.  For $p=-3$, the dimensionless spectrum $k^3P(k)$ is scale invariant, whereas for $p=0$ it scales as $k^3$. 
The broader scan varied the equation-of-state parameter
$\w=0$ or $1/3$, the fluctuation power $p$, the box infrared scale
$k_{\rm IR}$, the initialized UV cutoff $k_{\rm UV}$, the quartic coupling
$\lambda$, and the random seed.  
We choose $k_{\rm UV}=\{0.25,1\}m_0$ and
$\lambda=\{0.25,1\}m_0^2/v^2$, so that the
$N_{\rm lat}^3=256^3$ lattice resolves the initial monopole-core scale
$1/(\sqrt{\lambda}v)$.

For each simulation run, we identify global monopoles by the winding number of
the normalized scalar field on lattice cubes.  Our primary abundance estimator
is a clustered cube-flow count per Hubble volume.  We first apply the short
cube flow described in Appendix~\ref{app:measure}, group neighboring nonzero
flowed cubes into connected clusters, and assign each non-boundary-crossing
cluster to a Hubble patch (see also Appendix~\ref{app:energy_diag} for other measurements).  The count used in the fits is
\begin{align}
  \xi_{\rm count}
  =
  \left\langle N_{{\rm tot},H}\right\rangle_H,
  \qquad
  N_{{\rm tot},H}=N_{{\rm mon},H}+N_{{\rm anti},H}.
\end{align}
Here $\langle\cdots\rangle_H$ denotes the average over complete Hubble patches.

\paragraph{Fat Monopole Analysis}
We use a fat-monopole prescription analogous to the Press--Ryden--Spergel
prescription for cosmic strings~\cite{Press:1989yh}.  Appendix~\ref{app:constant_lambda}
shows a separate fixed-$\lambda$ check.  The vacuum expectation
value is kept fixed, while the quartic coupling is made time dependent,
\begin{align}
  \lambda_{\rm eff}(a)=\frac{\lambda}{a^2}.
\end{align}
We denote the corresponding physical radial mass by
$m_r(a)\simeq \sqrt{2\lambda_{\rm eff}(a)}\,v$.  In the fat-monopole
prescription $m_r\propto a^{-1}$, and the comoving monopole core width remains
approximately constant.  This allows us to keep the monopole cores resolved
for a long time on a fixed comoving lattice, giving the network time to relax
toward its late-time scaling regime.

This prescription differs from the physical fixed-$\lambda$ theory (see Appendix \ref{app:constant_lambda}), in which
the physical core width is fixed and the comoving core width shrinks as
$a^{-1}$.  However, for global monopoles the total energy is dominated by the
long-range Nambu--Goldstone gradient field rather than by the core.  We
therefore use the fat-monopole prescription as the main numerical setup for
studying the large-scale scaling of the monopole network.  The quantities
$\xi_{\rm count}$ are then compared across box sizes, initial UV cutoffs, and random seeds to assess the
stability of the scaling behavior. 
We also performed simulations with $N_{\rm lat}=512$ and $896$ as consistency checks.

Another feature of this fat monopole setup is that the radial-mode energy does not dominate the system.  For coherent oscillations around the minimum,
% the effective
%potential is $m_{r}^2\delta h^2/(2a^2)$, where
%$\delta h=(|\f|-v)$.  The adiabatic invariant then implies that the radial
energy density of the radial modes redshifts approximately as radiation.  The slowly varying component
which controls the monopole abundance is therefore the long-range
Nambu--Goldstone gradient energy in the radiation and matter dominated backgrounds. 
This is the reason that the total-energy scaling can be used as an auxiliary
diagnostic of monopole scaling, as discussed in Appendix~\ref{app:energy_diag}. \\

We run ensembles in both matter- and radiation-dominated backgrounds with
the conformal time in the
range $\eta=[1,100]/m_0$. This corresponds to cosmic-time dynamic ranges of $\O(10^4) \AND \O(10^6)$ in radiation- and matter-dominated backgrounds, respectively, which are orders of magnitude longer than previous numerical studies.

\subsection{Numerical results}

\begin{figure}[t]
\centering
\subfloat[$\w=0$]{
  \includegraphics[width=0.48\textwidth]{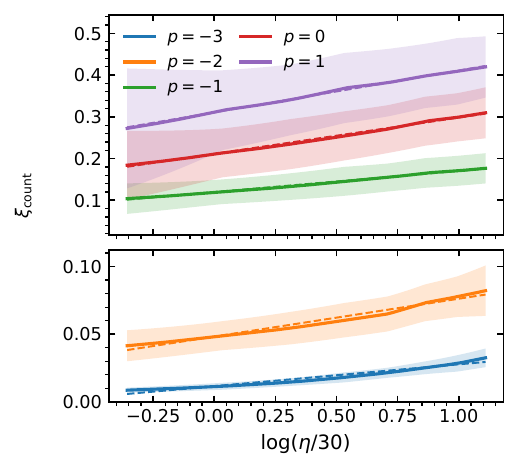}}
\hfill
\subfloat[$\w=1/3$]{
  \includegraphics[width=0.48\textwidth]{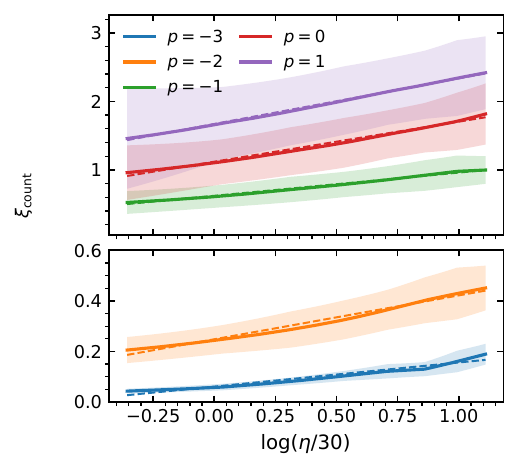}}
\caption{Late-time evolution of the clustered cube-flow monopole count in the
fiducial sample.  The curves show the unweighted mean over the characteristic
$(k_{\rm UV},\lambda)$ group-mean curves for each initial spectral tilt $p$.
The shaded bands show the standard deviation among those group-mean curves
after interpolation to a common conformal-time grid. In both panels, split vertical ranges
keep both the low- and high-abundance branches visible. 
Dashed lines are the corresponding fits with \Eq{Bfit}.}
\label{fig:xi_scaling}
\end{figure}

We first inspect the time evolution directly.  For each pair of
$(\w,p)$, where $\w$ is the background equation-of-state parameter and $p$ is
the spectral tilt defined above, we reconstruct the clustered count
$\x_{\rm count}(\y)$ for every usable run.  Figure~\ref{fig:xi_scaling} shows
the corresponding group-mean scaling curves.  The late-time behavior is not a
horizontal band.  For all spectra in the fiducial sample the curves clearly show 
positive drifts, and the drift is considerably larger in the radiation-dominated
background than in the matter-dominated background when it is measured as an
additive slope in $\x$.

To quantify the drift without assuming a detailed microscopic model, we fit
each run at late times to the pivoted logarithmic form
\beq
  \x_{\rm count}(\y)=A_{30}+B\log\frac{\y}{30},
  \qquad \y\gtrsim 20 .
  \label{eq:Bfit}
\eeq
The slope $B$ is unchanged by this choice of pivot, while $A_{30}$ is the fitted
normalization of $\x_{\rm count}$ at $\y=30$. 
 Fits with $\y>20$ and $\y>30$ give the same qualitative ordering and compatible
fractional slopes $\gamma$, as defined below.
In the fiducial sample used below, the group-mean slopes are positive for all five initial spectra in both backgrounds.  We therefore use the fitted slope
$B$ itself as the primary diagnostic of the logarithmic drift.

The averaged values obtained from the clustered cube-flow monopole count are summarized
in Table~\ref{tab:Bfits}.  For each fixed $(\w,p,k_{\rm UV},\lambda)$ we first
average over the available random seeds and $k_{\rm IR}$ choices.  The quoted
central value is then the unweighted mean over the resulting
$(k_{\rm UV},\lambda)$ group means, and the quoted uncertainty is the standard
deviation of those group means.  

As auxiliary diagnostics, we also compare the logarithmic fit with a constant
fit using the AIC difference and check run-level SEM/bootstrap errors within
fixed-input ensembles.  For the clustered-count fits, the group-averaged AIC
difference favors the logarithmic fit for all five values of $p$ in both
cosmological backgrounds, with $\Delta{\rm AIC}\simeq 18$--28. The run-level statistical
errors are smaller than the group-to-group scatter and are therefore not used
as the main error bars.

\begin{table}[t]
\centering
\caption{Late-time logarithmic fits for the $\eta\gtrsim20$ window in the
fiducial sample.  The fit is
$\xi_{\rm count}=A_{30}+B\log(\eta/30)$.  Here $\xi_{\rm count}$ is the clustered cube-flow
Hubble-patch count, and
$\gamma=d\log\xi_{\rm count}/d\log(m_r/H)$.  The central values are unweighted
means over the characteristic $(k_{\rm UV},\lambda)$ groups in each
$(\omega,p)$ class.  The errors are the standard deviations among those group
means.}
\label{tab:Bfits}
\small
\setlength{\tabcolsep}{4.5pt}
\renewcommand{\arraystretch}{1.18}
\arrayrulecolor{black!55}
\begin{tabular}{c!{\vrule width 0.7pt}ccc!{\vrule width 0.9pt}ccc}
\toprule
\rowcolor{black!8}
& \multicolumn{3}{c!{\vrule width 0.9pt}}{$\w=0$}
& \multicolumn{3}{c}{$\w=1/3$} \\
\rowcolor{black!8}
$p$
& \cellcolor{blue!8}$A_{30}$
& \cellcolor{blue!8}$B$
& \cellcolor{blue!8}$\g$
& \cellcolor{red!8}$A_{30}$
& \cellcolor{red!8}$B$
& \cellcolor{red!8}$\g$ \\
\midrule
\rowcolor{black!2}
$-3$
& \cellcolor{blue!4}$0.012\pm0.002$
& \cellcolor{blue!4}$0.016\pm0.003$
& \cellcolor{blue!4}$0.907\pm0.028$
& \cellcolor{red!4}$0.061\pm0.012$
& \cellcolor{red!4}$0.095\pm0.023$
& \cellcolor{red!4}$0.992\pm0.027$ \\
$-2$
& \cellcolor{blue!4}$0.048\pm0.011$
& \cellcolor{blue!4}$0.028\pm0.005$
& \cellcolor{blue!4}$0.472\pm0.045$
& \cellcolor{red!4}$0.248\pm0.060$
& \cellcolor{red!4}$0.174\pm0.035$
& \cellcolor{red!4}$0.552\pm0.048$ \\
\rowcolor{black!2}
$-1$
& \cellcolor{blue!4}$0.120\pm0.032$
& \cellcolor{blue!4}$0.051\pm0.011$
& \cellcolor{blue!4}$0.382\pm0.111$
& \cellcolor{red!4}$0.623\pm0.167$
& \cellcolor{red!4}$0.338\pm0.065$
& \cellcolor{red!4}$0.465\pm0.092$ \\
$0$
& \cellcolor{blue!4}$0.211\pm0.065$
& \cellcolor{blue!4}$0.087\pm0.029$
& \cellcolor{blue!4}$0.393\pm0.194$
& \cellcolor{red!4}$1.120\pm0.345$
& \cellcolor{red!4}$0.584\pm0.148$
& \cellcolor{red!4}$0.462\pm0.161$ \\
\rowcolor{black!2}
$1$
& \cellcolor{blue!4}$0.311\pm0.108$
& \cellcolor{blue!4}$0.100\pm0.050$
& \cellcolor{blue!4}$0.349\pm0.246$
& \cellcolor{red!4}$1.672\pm0.579$
& \cellcolor{red!4}$0.664\pm0.240$
& \cellcolor{red!4}$0.396\pm0.222$ \\
\bottomrule
\end{tabular}
\arrayrulecolor{black}
\end{table}

The additive slope is positive in both backgrounds for all five spectra in the
fiducial sample.  The radiation-dominated slope is larger by a factor
$B_{\rm RD}/B_{\rm MD}\simeq5.9$--$6.7$ when the comparison is made using $B$.
This ratio, however, should not be interpreted by itself as a universal
physical enhancement.
The reason is that $B$ measures an additive change of $\x$ per logarithmic
interval of conformal time, and may therefore be sensitive to the normalization and
the residual curvature of $\x(\y)$ in each background.  In particular, in a
setup with a time-dependent radial mass, the same conformal time does not
necessarily correspond to the same physical stage of monopole formation and
network relaxation.

\begin{figure}[t]
\centering
\subfloat[$B_{\rm RD}/B_{\rm MD}$]{
  \includegraphics[width=0.32\textwidth]{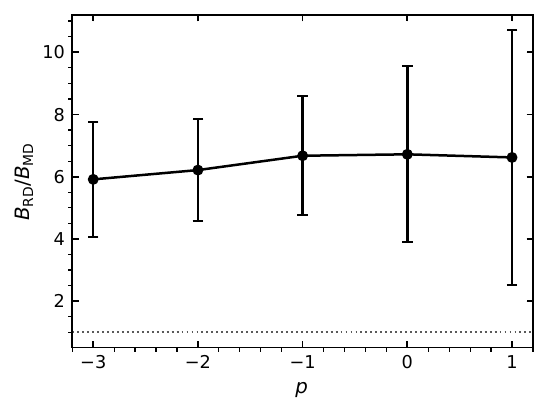}}
\hfill
\subfloat[$d\log\x_{\rm count}/d\log(m_r/H)$]{
  \includegraphics[width=0.32\textwidth]{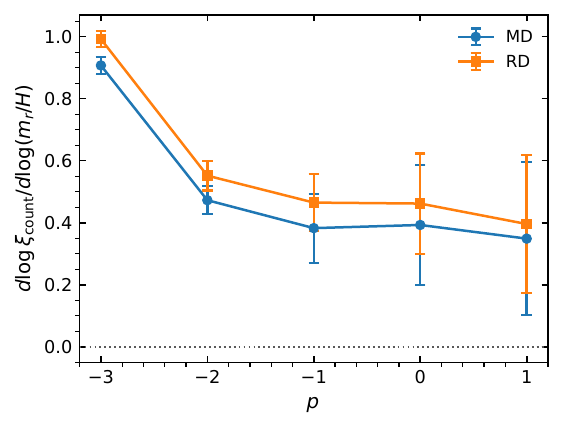}}
\hfill
\subfloat[$\g_{\rm RD}/\g_{\rm MD}$]{
  \includegraphics[width=0.32\textwidth]{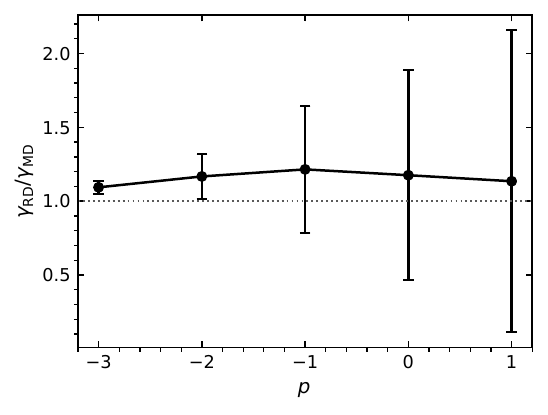}}
\caption{Summary of the late-time logarithmic fits in the fiducial sample.  The
left panel shows the radiation-to-matter ratio obtained from the additive
slope $B$ in Eq.~\eqref{eq:Bfit}.  The middle and right panels instead use the
fractional response to the physical hierarchy $m_r/H$, which removes most of
the apparent background dependence.  The plotted error bars are standard
deviations among the $(k_{\rm UV},\lambda)$ group means.}
\label{fig:fit_summary}
\end{figure}

A useful way to compare the drift in different backgrounds is to use the
fractional response\footnote{
The approximate stability of $\g$ against changes in the initial conditions and
the equation of state may suggest that $\x$ behaves as an effective power law,
$\x \propto (m_r/H)^{\g}$, over certain time interval.
}
\beq
  \g
  \equiv
  \frac{d\log \x_{\rm count}}{d\log(m_r/H)} .
  \label{eq:gamma_def}
\eeq
Note that in the fat-monopole prescription $m_r\propto a^{-1}$. For a constant
equation-of-state background with $w\neq -1/3$, one has
$m_r/H\propto\eta$.
$\g$ measures
the growth of $\x_{\rm count}$ relative to its normalization. 
This is useful
because the additive slope $B$ is sensitive to the overall abundance, which
differs substantially between matter and radiation domination. The resulting
values of $\gamma$ are listed together with $A_{30}$ and $B$ in
Table~\ref{tab:Bfits}.  The comparison between the additive and fractional
measures is shown in Fig.~\ref{fig:fit_summary}.

Energy-based abundance diagnostics and their logarithmic fit coefficients are collected in Appendix~\ref{app:energy_diag}.  A separate fixed-$\lambda$ run is shown in Appendix~\ref{app:constant_lambda} as a qualitative check of the fat-core prescription.

The numerical results support the following picture.  The global monopole
network does not settle rapidly to a constant value of
$\x_{\rm count}$.  Instead, over the simulated late-time interval, it retains
a positive logarithmic evolution.  The additive slope $B$ is much larger in
radiation domination, but the fractional response
$\g=d\log\x_{\rm count}/d\log(m_r/H)$ is similar in the two backgrounds.
The measured $\g$ is far less sensitive to the background equation of state
than $B$, and is also relatively stable against changes in the initial
conditions for blue-tilted power spectra.
 Given the long simulation times and the persistence of these
positive logarithmic corrections across a range of initial conditions, these
results strongly suggest that the commonly assumed constant-scaling ansatz is
inadequate and should be revisited.

\section{Conclusions and Discussion}

We have simulated global monopole networks with several initial fluctuation
spectra in both matter- and radiation-dominated backgrounds.  The main result
is that the late-time monopole count is not consistent with an exactly constant
scaling parameter.  The fitted slope in Eq.~\eqref{eq:Bfit} is positive for all
initial spectra used in this work.  When expressed as the
fractional response
$\g=d\log\x_{\rm count}/d\log(m_r/H)$, the equation-of-state dependence, as well
as the initial-condition dependence for blue-tilted spectra, is weaker than it
appears in the additive slope $B$.  Consequently, our simulations indicate that
the monopole number density violates the usual constant-scaling assumption,
even though the energy of an isolated global monopole grows linearly, rather
than logarithmically, with the infrared cutoff.  This result may also provide
insight into the logarithmic scaling observed in cosmic string networks.

\subsection{Interpretation of the violation of constant scaling}

We now give a simple interpretation of the violation of constant scaling.
Let $n_M$ be the total monopole plus antimonopole number density.  Its coarse-grained
evolution can be written as~\cite{Yamaguchi:2001rf}
\beq
\dot{n}_M + 3H n_M = - \frac{n_M}{\tau_{\rm ann}},
\laq{stationary}
\eeq
where $\tau_{\rm ann}$ is the characteristic annihilation time.  If monopoles
move relativistically and annihilate after traveling a distance of order the
typical separation $D\simeq n_M^{-1/3}$, then
$\tau_{\rm ann}\sim D$.  The stationary solution is then
$n_M\sim H^3$, or an approximately constant number of monopoles per Hubble
volume.

Our simulations indicate that this estimate misses a slowly varying correction.
A simple way to parametrize the effect is to allow the annihilation time to
depend on the ratio of the monopole separation to the core size,
\begin{equation}
  \tau_{\rm ann}
  =
  D\left[
    c_0+c_1\log\frac{D}{r_{\rm core}}+\cdots
  \right],
  \qquad
  r_{\rm core}\simeq m_r^{-1}.
\end{equation}
Such a dependence can arise if a non-negligible fraction of monopole--antimonopole
pairs carry angular momentum or impact parameters large enough that the cores
do not collide on the first encounter.  Hubble friction and Goldstone radiation
then have to remove angular momentum before annihilation can occur efficiently,
producing a logarithmic correction.  Since
$D$ is of order $H^{-1}$ in the scaling regime, the logarithm becomes
$\log(D/r_{\rm core})\simeq\log(m_r/H)$, up to order-one factors and the weak
dependence on $\x$.  This argument explains why a logarithm of the hierarchy
between the radial scale and the Hubble scale can enter the annihilation
efficiency.
In a more concrete model of monopole annihilation, the measured $B$ and $\g$
in Table~\ref{tab:Bfits} can be used to determine the corresponding effective
parameters.

\subsection{Production of dark matter candidates}

The violation of constant scaling would affect various relevant phenomena,
such as the spectra of primordial black holes and gravitational waves; see
Ref.~\cite{Aburatani:2026rct}.  For instance, a positive logarithmic correction
to the scaling abundance implies that the population of magnetic primordial
black holes (PBHs), which are produced mostly near the end of the scaling
regime, would be enhanced compared with that of neutralized PBHs.  The
approximately scale-invariant gravitational-wave spectrum emitted during the
scaling regime in radiation domination may also become blue tilted.  

The production of axions, dark photons, and monopole dark matter is also
relevant in this context.  We briefly discuss these possibilities below, while
a more detailed study is presented elsewhere.  In particular, if dark photons
are produced from a global-like monopole network, the gauge coupling must be
sufficiently small so that the network retains the global-monopole-like
long-range dynamics before the end of scaling.  Such a small gauge coupling is,
however, also a generic requirement for dark-photon dark matter whose mass is
generated by the Higgs mechanism~\cite{Kitajima:2021bjq,Nakagawa:2022knn,East:2022rsi,Cyncynates:2023zwj,Cyncynates:2024yxm,Kitajima:2024jfl}.

\paragraph{Dark photons and axions from the monopole network}

The logarithmic growth of $\x$ is also relevant for dark-sector relics.  We
consider the case in which part of the $\O(3)$ symmetry is gauged or explicitly
broken, so that the global-monopole scaling regime terminates when the
corresponding screening mass or explicit-breaking scale becomes comparable to
the Hubble scale. 

Before the end of the scaling regime, the monopole energy is dominated by the
Goldstone gradient field for $m_r\gg H$.  We therefore assume
that all monopoles collapse and the energy remaining at the end of scaling is converted into
dark-sector particles with typical energy $E_{\rm DM}$. 
This scenario should be regarded as a benchmark estimate; the precise endpoint of dark-relic production, and whether monopoles remain, depend on the gauge coupling and/or explicit breaking terms, which will be studied elsewhere. 

We denote this relic
component by $\Omega_{\rm DP}$, where DP stands for the produced dark photons or pseudo Nambu-Goldstone bosons (axions). The monopole energy entering the estimate is
$M[R]\simeq4\pi v^2R$ evaluated at the inter-monopole separation
$R\simeq n_M^{-1/3}$ (one may instead use $\rho_{\rm tot}$ given in Appendix~\ref{app:energy_diag}). The resulting abundance is
\begin{align}
\Omega_{\rm DP}h^2
&=
\frac{s_0}{\rho_c h^{-2}}
m_{\rm DM}
\frac{n_M M[R]}{E_{\rm DM}s}
\\
&=
0.1
\frac{m_{\rm DM}}{2.5\times10^{-20}\EV}
\(\frac{\x}{100}\)^{2/3}
\(\frac{v}{2\times10^{15}\GEV}\)^2
\left.
\(\frac{T}{10\KEV}
\frac{H}{E_{\rm DM}}
g_*g_{*s}^{-1}
\)
\right|_{T=T_{\rm end}}.
\end{align}
Here the subscript ``end'' denotes the end of the scaling regime, $s_0$ is
the present entropy density, $\rho_c$ is the critical density, and
$g_*,g_{*s}$ are the usual energy and entropy relativistic degrees of freedom.
We require $m_{\rm DM}<H_{\rm end}$ if the produced particles are (semi)relativistic
at production with the typical energy $E_{\rm DM}\sim H$.  According to the equivalence theorem, this formula should work both for the axion and (longitudinal) dark photon case; see
Refs.~\cite{Jaeckel:2010ni,Ringwald:2012hr,Arias:2012az,Graham:2015ouw,Marsh:2015xka,Irastorza:2018dyq,DiLuzio:2020wdo,Albertus:2026fbe,Arza:2026rsl}
for reviews of light dark sectors.

The distinction from the dark matter production from an ordinary long-lived cosmic string network~\cite{ Long:2019lwl, Nakayama:2021avl, Kitajima:2022lre} is that depending on the mass effects the
monopole scaling regime can terminate when the long-range force is screened.
If the network instead remains active for a long time, its radiation and
stress-energy can be constrained by CMB and gravitational-wave observables
\cite{Lopez-Eiguren:2017dmc}.
In the screened monopole scenario, most of the energy can be dumped into the
dark sector before such late-time signals become dominant.

\paragraph{Hidden Monopole and Dark Photon Co-Dark Matter}

One can also consider (hidden) monopole dark matter by fully gauging the $\O(3)$ symmetry ~\cite{Murayama:2009nj,Baek:2013dwa, Khoze:2014woa,Kawasaki:2015lpf}.\footnote{
Ref.~\cite{Murayama:2009nj} focused on the initial Kibble--Zurek abundance of
point-like topological defects and treated the defects essentially as stable
particle relics after production.  This is complementary to the present work,
where the subsequent scaling evolution driven by long-range would-be Goldstone
gradients determines the monopole abundance.
}
The long-range interaction is then screened, and monopole--antimonopole annihilation effectively ceases when
$
m_W=gv\sim H_{\rm end},
$
where $g$ is the hidden gauge coupling.  The resulting abundance is
\begin{align}
\Omega_m h^2
&=
\frac{s_0}{\rho_c h^{-2}}
\frac{\rho_H}{s}\Big|_{t=t_{\rm end}}
=
0.03
\(\frac{\x}{100}\)^{2/3}
\(\frac{v}{10^{11}\GEV}\)^2
\left.
\(\frac{T}{1\TEV}
g_*
g_{*s}^{-1}
\)
\right|_{T=T_{\rm end}},
\\
M_m
&=
3.4\times10^{10}{\rm gram}
\(\frac{v}{10^{11}\GEV}\)^2
\(\frac{\TEV}{T_{\rm end}}\)^2
\sqrt{\frac{106.75}{g_*(T_{\rm end})}}
\(\frac{\x}{100}\)^{-1/3}.
\end{align}

Here $\rho_H$ denotes the monopole energy density at the end of the
long-range scaling regime.  A comparable contribution to the dark matter
abundance can be obtained from massive charged gauge bosons produced during the scaling
regime.  A massless gauge boson component instead behaves as dark radiation and
is typically subdominant in the benchmark region considered here.
The charged gauge boson mass is
\beq
m_W
=
10^{-3}\EV
\(\frac{T}{\TEV}\)^2
\sqrt{\frac{g_*}{106.75}}.
\eeq
This scenario therefore naturally predicts a multi-component dark matter sector consisting of a heavy monopole component and a light gauge boson component.

\section*{Acknowledgements}
This work is supported by JSPS KAKENHI Grant Nos. 22K14029 (W.Y.), 23K22486 (W.Y.), and  26K00695 (W.Y.). W.Y. is also supported by Selective Research Fund and Incentive Research Fund from Tokyo Metropolitan University. 

\appendix 

\section{Hubble-patch observables}
\label{app:measure}

This appendix defines the Hubble-patch observables used in the analysis.  Our
primary monopole-abundance estimator is the clustered topological count
$\xi_{\rm count}$, while the energy-based quantities are used as auxiliary
diagnostics.

At each output time, we partition the simulation box into complete Hubble-sized
blocks.  The comoving Hubble length in lattice units is
\begin{equation}
  L_H^{\rm lat}
  =
  \frac{a/a'}{dx},
\end{equation}
where a prime denotes a derivative with respect to conformal time and $dx$ is
the lattice spacing.  We use only outputs for which $L_H^{\rm lat}$ is larger
than one lattice spacing.  The number of complete Hubble blocks along each
spatial direction is
\begin{equation}
  N_H
  =
  \left\lfloor
  \frac{N_{\rm lat}}{L_H^{\rm lat}}
  \right\rfloor ,
\end{equation}
where $N_{\rm lat}$ is the number of lattice cells per side.  Thus the total
number of complete Hubble blocks is
\begin{equation}
  N_{\rm block}=N_H^3 .
\end{equation}
The side length of the lattice region included in the Hubble-patch average is
\begin{equation}
  N_{\rm used}
  =
  \left\lfloor
  N_H L_H^{\rm lat}
  \right\rfloor .
\end{equation}
The remaining boundary cells are not used.  This avoids assigning partial
Hubble volumes to the Hubble-patch statistics.

We compute the monopole charge geometrically from the map of each elementary
cube to the vacuum manifold.  The normalized scalar direction is
\begin{equation}
  \hat n
  =
  \frac{\boldsymbol{\phi}}{|\boldsymbol{\phi}|}.
\end{equation}
For a triangular face with vertices $\hat n_1,\hat n_2,\hat n_3$, the oriented
solid angle on the unit sphere is evaluated as
\begin{equation}
  \Omega(\hat n_1,\hat n_2,\hat n_3)
  =
  2\tan^{-1}
  \frac{
    \hat n_1\cdot(\hat n_2\times \hat n_3)
  }{
    1+\hat n_1\cdot\hat n_2
     +\hat n_2\cdot\hat n_3
     +\hat n_3\cdot\hat n_1
  } .
\end{equation}
The winding charge of an elementary cube is then obtained by summing the
oriented areas of the twelve triangles on the cube boundary,
\begin{equation}
  Q_{\rm cube}
  =
  \frac{1}{4\pi}
  \sum_{\Delta\in\partial{\rm cube}}
  \Omega_\Delta .
\end{equation}
This is the standard lattice definition of the winding number based on the
image area on the vacuum manifold~\cite{Berg:1981er}.

For the monopole count used in the main text, we first apply a short local flow
to the eight normalized field directions on each cube.  This removes
lattice-scale angular noise while preserving the constraint
$|\hat n|=1$.  One flow step is
\begin{equation}
  \hat n_i
  \to
  {\cal N}\left[
  \hat n_i
  +
  \epsilon
  \left(
  \sum_{j\in{\rm n.n.}(i)}(\hat n_j-\hat n_i)
  -
  \hat n_i
  \hat n_i\cdot
  \sum_{j\in{\rm n.n.}(i)}(\hat n_j-\hat n_i)
  \right)
  \right],
\end{equation}
where ${\cal N}$ denotes normalization to unit length, and ${\rm n.n.}(i)$ is
the set of vertices connected to vertex $i$ by an edge of the cube.  In the
analysis we use four flow steps with $\epsilon=0.02$.

After the flow, each cube is assigned the integer charge
${\rm round}(Q_{\rm cube}^{\rm flow})$.  Cubes with nonzero rounded charge are
then grouped into nearest-neighbor connected clusters.  The charge of a
cluster $C$ is
\begin{equation}
  Q_C
  =
  \sum_{{\rm cubes}\in C}
  {\rm round}(Q_{\rm cube}^{\rm flow}) .
\end{equation}
A charged cluster that crosses a Hubble-block boundary is excluded from the
single-patch assignment.  Otherwise, the cluster is assigned to the Hubble block
that contains it.  We then define
\begin{equation}
  N_{{\rm mon},H}
  =
  \sum_{C\subset H}
  \max(Q_C,0),
  \qquad
  N_{{\rm anti},H}
  =
  \sum_{C\subset H}
  \max(-Q_C,0),
\end{equation}
and
\begin{equation}
  N_{{\rm tot},H}
  =
  N_{{\rm mon},H}
  +
  N_{{\rm anti},H}.
\end{equation}
The abundance estimator used in the main analysis is
\begin{equation}
  \xi_{\rm count}
  =
  \left\langle N_{{\rm tot},H}\right\rangle_H .
\end{equation}
The simulation output also contains the direct global cube-flow sum over
elementary cubes.  We use this quantity only as a cross-check, because it does
not merge neighboring charged cubes that belong to the same monopole core.

We also compute energy-based Hubble-patch observables.  For each complete
Hubble block, the gradient and total scalar energy densities are
\begin{equation}
  \rho_{{\rm grad},H}
  =
  \frac{1}{N_{{\rm site},H}}
  \sum_{\boldsymbol{x}\in H}
  \rho_{\rm grad}(\boldsymbol{x}),
\end{equation}
and
\begin{equation}
  \rho_{{\rm tot},H}
  =
  \frac{1}{N_{{\rm site},H}}
  \sum_{\boldsymbol{x}\in H}
  \left[
    \rho_{\rm kin}(\boldsymbol{x})
    +
    \rho_{\rm grad}(\boldsymbol{x})
    +
    \rho_{\rm pot}(\boldsymbol{x})
  \right].
\end{equation}
To characterize patch-to-patch fluctuations, we normalize these quantities by
their Hubble-patch averages at the same output time,
\begin{equation}
  x_{{\rm grad},H}
  =
  \frac{\rho_{{\rm grad},H}}
       {\langle \rho_{{\rm grad},H} \rangle_H},
  \qquad
  x_{{\rm tot},H}
  =
  \frac{\rho_{{\rm tot},H}}
       {\langle \rho_{{\rm tot},H} \rangle_H}.
\end{equation}
Here $\langle\cdots\rangle_H$ denotes an average over all complete Hubble
patches at fixed output time.

\section{Energy-based diagnostics}
\label{app:energy_diag}

This appendix collects the energy-based checks that are not used as the primary
monopole count in the main text.  We define
\begin{equation}
  \x_{\rm grad}
  =
  \frac{\langle\rho_{{\rm grad},H}\rangle_H}{4\pi v^2H^2},
  \qquad
  \x_{\rm totE}
  =
  \frac{\langle\rho_{{\rm tot},H}\rangle_H}{4\pi v^2H^2}.
  \label{eq:energy_xi}
\end{equation}
The gradient estimator has the expected normalization for a global monopole
whose energy is dominated by the Goldstone gradient field.  The total-energy
estimator is useful as a diagnostic, but it is not a clean monopole counter:
kinetic energy, potential energy, and residual scalar radiation are included in
the numerator.

The same logarithmic fits used in the main text are applied to these quantities,
\begin{equation}
  \x_X(\y)=A_{30,X}+B_X\log\frac{\y}{30},
  \qquad \y\gtrsim 20,
  \qquad
  \gamma_X=\frac{d\log\x_X}{d\log(m_r/H)},
  \qquad X=\{{\rm grad},{\rm totE}\}.
\end{equation}
The resulting coefficients are shown in Tables~\ref{tab:gradlogfits}
and~\ref{tab:totelogfits}.  The gradient-energy slopes are broadly comparable
to the cube-flow slopes.  The total-energy slopes, especially in the
radiation-dominated runs, are substantially larger; this is consistent with
the interpretation that $\x_{\rm totE}$ is sensitive to non-monopole energy
stored in waves and radial motion.

Note that, in the absence of logarithmic corrections, the total energy in our
fat-monopole setup is expected to redshift as radiation during radiation
domination.  The reason is that the radial-mode oscillations redshift as
radiation in this prescription, while the energy density of a constant-scaling
monopole network also scales as the background radiation energy density. Thus, the total-energy scaling by itself already indicates a departure from constant scaling of the monopole network.

We have also checked that the total energy becomes radiation-like when the initial
field value is shifted away from zero $\vev{\f_a}\neq0$.

\begin{figure}[p]
\centering
\includegraphics[width=0.85\textwidth]{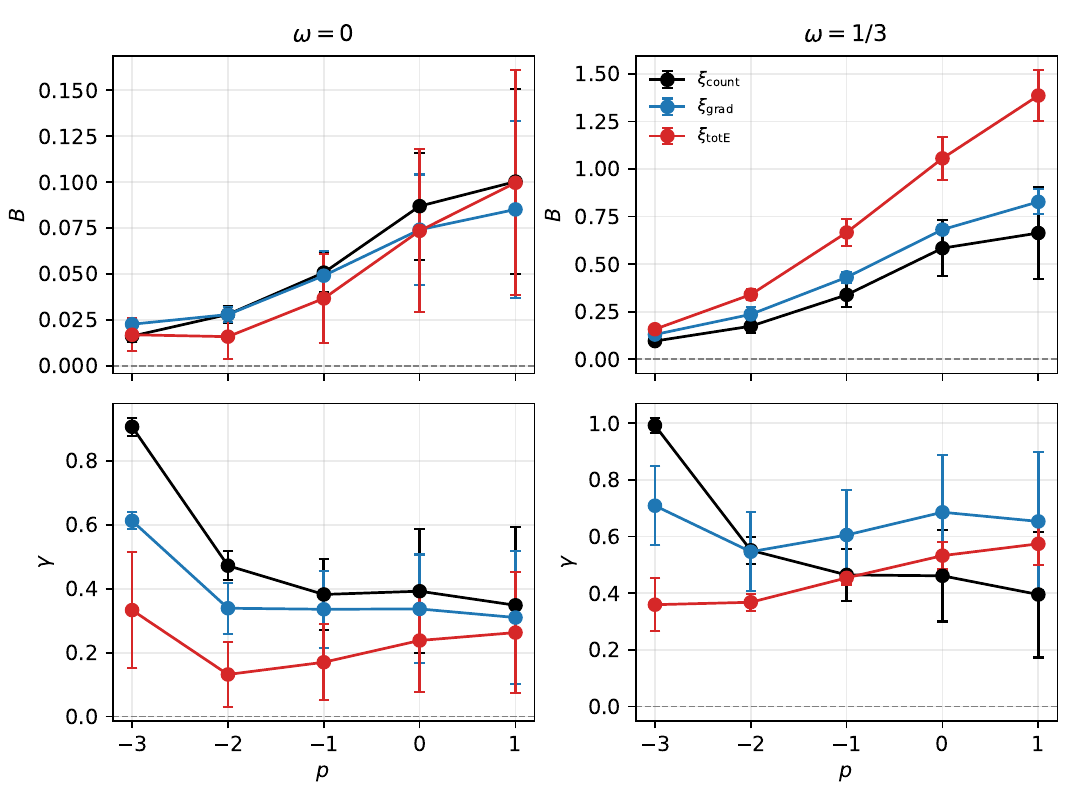}
\caption{Late-time logarithmic coefficients for the clustered cube-flow count, the
gradient-energy estimator, and the total-energy estimator.  The upper panels
show $B_X$ in the fit $\x_X=A_{30,X}+B_X\log(\y/30)$, while the lower panels show
$\gamma_X=d\log\x_X/d\log(m_r/H)$.  Error bars are standard deviations among
the $(k_{\rm UV},\lambda)$ group means in the fiducial sample.}
\label{fig:energy_coefficients_appendix}
\end{figure}

\begin{figure}[p]
\centering
\includegraphics[width=0.85\textwidth]{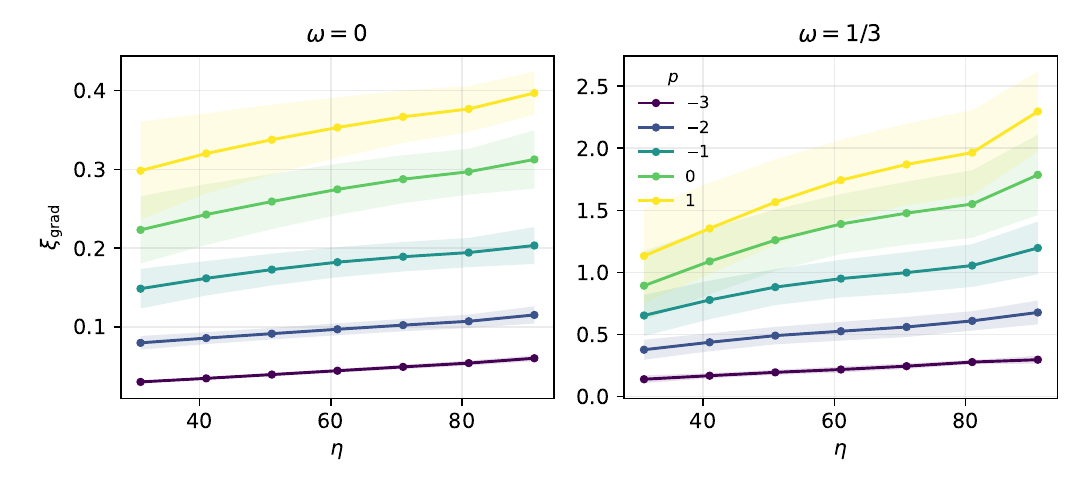}
\caption{Time evolution of the gradient-energy estimator $\x_{\rm grad}$.
The curves are means over the fiducial $(k_{\rm UV},\lambda)$ group means; shaded bands
show the corresponding group standard deviations.  The left and right panels
show matter and radiation domination, respectively.}
\label{fig:xigrad_scaling_appendix}
\end{figure}

\begin{figure}[p]
\centering
\includegraphics[width=0.85\textwidth]{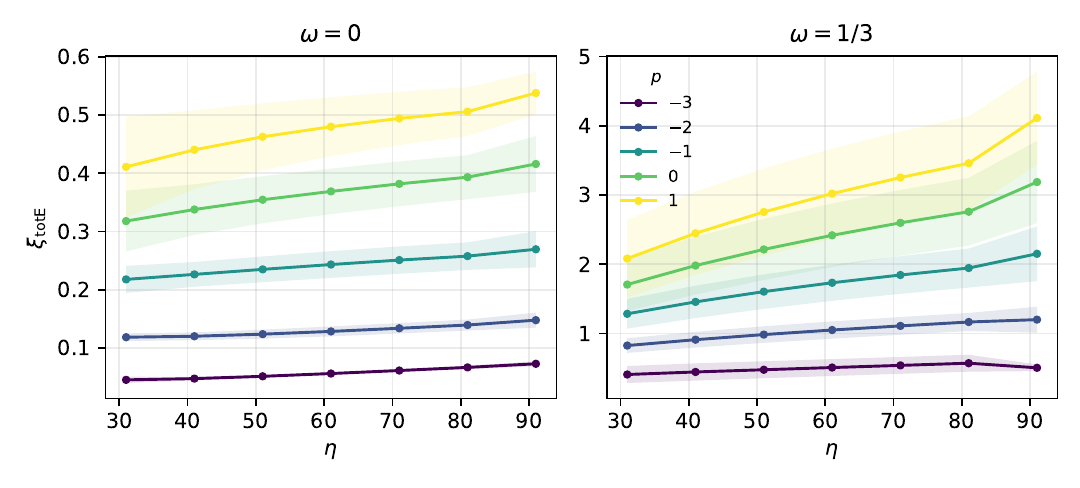}
\caption{Time evolution of the total-energy estimator $\x_{\rm totE}$, with the
same grouping convention as Fig.~\ref{fig:xigrad_scaling_appendix}.  This
quantity is not used as the primary monopole count because it includes
non-monopole scalar energy.}
\label{fig:xitote_scaling_appendix}
\end{figure}

\begin{table}[p]
\centering
\caption{Late-time logarithmic fits for the gradient-energy estimator
$\x_{\rm grad}=A_{30,\rm grad}+B_{\rm grad}\log(\y/30)$ and
$\gamma_{\rm grad}=d\log\x_{\rm grad}/d\log(m_r/H)$.  Errors are standard
deviations among the $(k_{\rm UV},\lambda)$ group means in the fiducial sample.}
\label{tab:gradlogfits}
\scriptsize
\setlength{\tabcolsep}{3.2pt}
\begin{tabular}{c|ccc|ccc}
\toprule
& \multicolumn{3}{c|}{$\w=0$} & \multicolumn{3}{c}{$\w=1/3$} \\
$p$ & $A_{30,\rm grad}$ & $B_{\rm grad}$ & $\gamma_{\rm grad}$
    & $A_{30,\rm grad}$ & $B_{\rm grad}$ & $\gamma_{\rm grad}$ \\
\midrule
$-3$ & $0.030\pm0.002$ & $0.023\pm0.002$ & $0.613\pm0.027$ & $0.140\pm0.025$ & $0.130\pm0.016$ & $0.709\pm0.139$ \\
$-2$ & $0.079\pm0.010$ & $0.028\pm0.004$ & $0.339\pm0.080$ & $0.371\pm0.077$ & $0.236\pm0.037$ & $0.547\pm0.139$ \\
$-1$ & $0.146\pm0.027$ & $0.049\pm0.013$ & $0.336\pm0.120$ & $0.638\pm0.169$ & $0.431\pm0.029$ & $0.606\pm0.160$ \\
$0$ & $0.222\pm0.045$ & $0.074\pm0.030$ & $0.337\pm0.170$ & $0.885\pm0.279$ & $0.682\pm0.025$ & $0.686\pm0.201$ \\
$1$ & $0.292\pm0.065$ & $0.085\pm0.048$ & $0.310\pm0.209$ & $1.137\pm0.378$ & $0.828\pm0.066$ & $0.654\pm0.243$ \\
\bottomrule
\end{tabular}
\end{table}

\begin{table}[p]
\centering
\caption{Late-time logarithmic fits for the total-energy estimator
$\x_{\rm totE}=A_{30,\rm totE}+B_{\rm totE}\log(\y/30)$ and
$\gamma_{\rm totE}=d\log\x_{\rm totE}/d\log(m_r/H)$.  The same grouping and
error convention as Table~\ref{tab:gradlogfits} is used.}
\label{tab:totelogfits}
\scriptsize
\setlength{\tabcolsep}{3.2pt}
\begin{tabular}{c|ccc|ccc}
\toprule
& \multicolumn{3}{c|}{$\w=0$} & \multicolumn{3}{c}{$\w=1/3$} \\
$p$ & $A_{30,\rm totE}$ & $B_{\rm totE}$ & $\gamma_{\rm totE}$
    & $A_{30,\rm totE}$ & $B_{\rm totE}$ & $\gamma_{\rm totE}$ \\
\midrule
$-3$ & $0.047\pm0.007$ & $0.017\pm0.009$ & $0.333\pm0.182$ & $0.406\pm0.122$ & $0.158\pm0.009$ & $0.360\pm0.094$ \\
$-2$ & $0.121\pm0.007$ & $0.016\pm0.012$ & $0.132\pm0.101$ & $0.820\pm0.108$ & $0.340\pm0.029$ & $0.368\pm0.030$ \\
$-1$ & $0.219\pm0.024$ & $0.037\pm0.024$ & $0.170\pm0.120$ & $1.272\pm0.213$ & $0.667\pm0.069$ & $0.454\pm0.024$ \\
$0$ & $0.318\pm0.054$ & $0.073\pm0.044$ & $0.238\pm0.162$ & $1.689\pm0.379$ & $1.056\pm0.114$ & $0.533\pm0.049$ \\
$1$ & $0.408\pm0.087$ & $0.100\pm0.061$ & $0.263\pm0.189$ & $2.058\pm0.551$ & $1.387\pm0.132$ & $0.575\pm0.074$ \\
\bottomrule
\end{tabular}
\end{table}

\begin{table}[p]
\centering
\caption{Late-time Hubble-normalized abundance estimates at $\y\simeq91$ in the
fiducial sample.  $\x_{\rm count}$ is the clustered cube-flow topological count,
$\x_{\rm grad}$ is the gradient-energy estimate in Eq.~\eqref{eq:energy_xi},
and $\x_{\rm totE}$ uses the total scalar energy with the same normalization.
The central values and errors are the means and standard deviations of the
available $(k_{\rm UV},\lambda)$ group means.}
\label{tab:energyxi}
\scriptsize
\setlength{\tabcolsep}{3.2pt}
\begin{tabular}{c|ccc|ccc}
\toprule
& \multicolumn{3}{c|}{$\w=0$} & \multicolumn{3}{c}{$\w=1/3$} \\
$p$ & $\x_{\rm count}$ & $\x_{\rm grad}$ & $\x_{\rm totE}$
    & $\x_{\rm count}$ & $\x_{\rm grad}$ & $\x_{\rm totE}$ \\
\midrule
$-3$ & $0.033\pm0.007$ & $0.060\pm0.004$ & $0.073\pm0.004$ & $0.19\pm0.04$ & $0.30\pm0.03$ & $0.50\pm0.05$ \\
$-2$ & $0.082\pm0.019$ & $0.12\pm0.01$ & $0.15\pm0.01$ & $0.45\pm0.09$ & $0.68\pm0.10$ & $1.20\pm0.19$ \\
$-1$ & $0.18\pm0.04$ & $0.20\pm0.02$ & $0.27\pm0.03$ & $1.00\pm0.20$ & $1.20\pm0.21$ & $2.15\pm0.39$ \\
$0$ & $0.31\pm0.06$ & $0.31\pm0.04$ & $0.42\pm0.05$ & $1.82\pm0.45$ & $1.79\pm0.32$ & $3.19\pm0.59$ \\
$1$ & $0.42\pm0.07$ & $0.40\pm0.03$ & $0.54\pm0.04$ & $2.42\pm0.53$ & $2.30\pm0.32$ & $4.11\pm0.67$ \\
\bottomrule
\end{tabular}
\end{table}

\section{Fixed-lambda check}

\label{app:constant_lambda}

The main numerical scan uses the fat-core prescription
$\lambda_{\rm eff}=\lambda/a^2$, for which the monopole core has an
approximately fixed comoving width.  As a complementary check, we also ran a
separate matter-dominated simulation with fixed quartic coupling,
$\lambda_{\rm eff}=\lambda$.  In this physical-core prescription the comoving
core shrinks with time, so the controlled time interval is shorter and the run
is not included in the ensemble averages quoted in the main text.  Its
spectral tilt is $p=0$ in the notation used above.

\begin{figure}[p]
\centering
\includegraphics[width=0.85\textwidth]{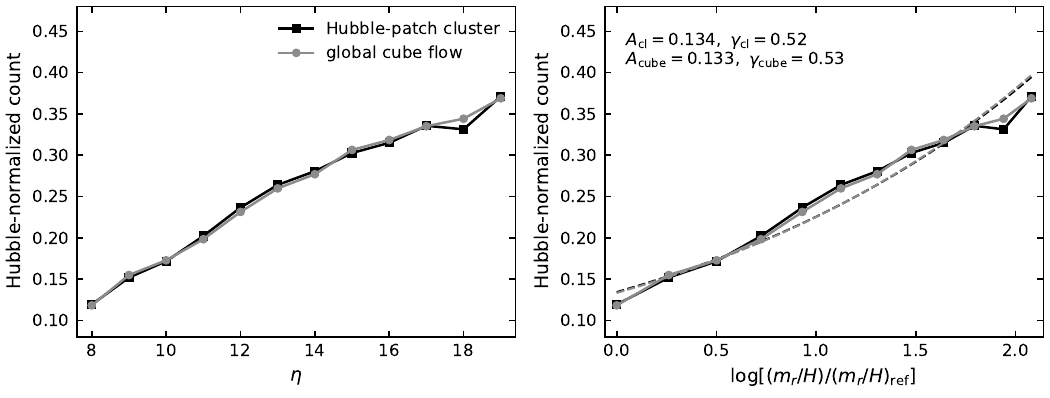}
\caption{Fixed-$\lambda$ check in a separate matter-dominated run.  Left:
Hubble-normalized monopole counts as functions of conformal time.  Right: the
same data plotted against $\log[(m_r/H)/(m_r/H)_{\rm ref}]$.  Dashed curves show
fits to $\log\xi=\log A+\gamma\log[(m_r/H)/(m_r/H)_{\rm ref}]$.  The black
curve uses the Hubble-patch clustered cube-flow definition used in the main
analysis, while the gray curve shows the direct global cube-flow count.}
\label{fig:constant_lambda_check}
\end{figure}

Figure~\ref{fig:constant_lambda_check} shows the Hubble-normalized monopole
count in this fixed-$\lambda$ run.  The black curve uses the same
Hubble-patch clustered cube-flow definition as the main analysis, while the
gray curve shows the direct global cube-flow count as a normalization
cross-check.  In the fixed-$\lambda$ prescription the physical radial mass is
constant, so the appropriate hierarchy variable is $m_r/H\propto H^{-1}$.  We
therefore fit
\begin{equation}
  \log\xi = \log A
  +\gamma\log\left[\frac{(m_r/H)}{(m_r/H)_{\rm ref}}\right] .
\end{equation}
The clustered Hubble-patch count gives $A_{\rm cl}\simeq0.134$ and
$\gamma_{\rm cl}\simeq0.52$.  This is somewhat larger than, but still
consistent at the order-one level with, the matter-dominated $p=0$ value
$\gamma=0.393\pm0.194$ in Table~\ref{tab:Bfits}.  The direct global cube-flow
count gives $A_{\rm cube}\simeq0.133$ and $\gamma_{\rm cube}\simeq0.53$.  
Because this single run covers a shorter controlled interval, we use it only
as a consistency check rather than as part of the main ensemble average.

\clearpage
\bibliography{scaling.bib}
\end{document}